\documentclass[12pt,preprint]{aastex}

\shorttitle{Asymmetry of Ly$\alpha$ scattering cross section}

\shortauthors{Lee}

\begin{document}
\title{Asymmetric Deviation of the Scattering Cross Section around 
Ly$\alpha$ by Atomic Hydrogen}

\author{Hee-Won Lee} 
\affil{Department of Astronomy and Space Sciences, Sejong University,
Seoul, 143-747, Korea \\
{\rm hwlee@sejong.ac.kr}}

\begin{abstract}
We investigate the asymmetry of the scattering cross section of radiation 
around Ly$\alpha$ by atomic hydrogen, which may be applied to analyses of 
scattering media with high column neutral hydrogen densities including
damped Ly$\alpha$ absorption systems of quasars.
The exact scattering cross section is given by the Kramers-Heisenberg formula 
obtained from the fully quantum mechanical second-order time dependent theory, 
where, in the case of hydrogen, each matrix element is given in a closed 
analytical form. The asymmetric deviation of the scattering cross section 
from the Lorentzian near the line center is computed by expanding the
Kramers-Heisenberg formula in terms of $\Delta\omega /\omega_{Ly\alpha}$,
where $\omega_{Ly\alpha}$ is the angular frequency of the Ly$\alpha$ 
transition and $\Delta\omega$ is the deviation of incident radiation 
from $\omega_{Ly\alpha}$.  To the first order of 
$\Delta\omega/\omega_{Ly\alpha}$, we obtain $\sigma(\omega) = 
\sigma_T(0.5 f_{12}
\omega_{Ly\alpha}/\Delta\omega)^2 (1-1.79\Delta\omega/\omega_{Ly\alpha})$, 
where $\sigma_T$ is the Thomson scattering cross section and $f_{12}=0.4162$ 
is the oscillator strength for the Ly$\alpha$ transition.
With this deviation, the line center of the damped wing profile apparently
shifts blueward of the true Ly$\alpha$ line center. In the case of a damped 
Ly$\alpha$ system with a H~I column density $5\times 10^{21}{\rm\ cm^{-2}}$, 
the apparent line center shift relative to the true center amounts 
to $0.2{\rm\ \AA}$ resulting in an underestimation of redshift
by $\Delta z\sim 10^{-4}$. A measurable underestimation by an amount of 
$\Delta z\sim 10^{-3}$ is expected for absorbing systems 
with $N_{HI}\ge 4\times 10^{22}{\rm\ cm^{-2}}$.  

\end{abstract}

\keywords{atomic physics --- radiative transfer --- cosmology --- quasar}

\section{Introduction}

The scattering process of electromagnetic radiation by an atom
is regarded as a second-order quantum mechanical process involving 
the annihilation of an incident photon
and the creation of a scattered photon.
Therefore, from the second-order time dependent perturbation theory,
we can obtain the scattering cross section, which is known as the
Kramers-Heisenberg formula (e.g. Sakurai 1969). The exact scattering cross 
section can be obtained after summing all bound state contributions
and integrating over all continuum state contributions 
related with the dipole processes of annihilation and creation of photons.
Hydrogen being a single electron atom, the energy eigenstate wavefunctions 
of hydrogen are analytically known and hence each matrix element appearing 
in the Kramers-Heisenberg formula is given in a closed form. 

Near resonance, the scattering cross section is excellently approximated by
the Lorentzian characterized by the narrow radiation damping term 
that can be interpreted as the life time of the excited state. The 
Lorentzian dependence is obtained since the cross section is dominated 
by the single contribution from the excited state relevant to the
resonance whereas the contributions from the other states are almost 
completely negligible. This Lorentzian approximation is quite generally
used in the astronomical community for analyses of resonance lines.

Far from resonance, the radiation damping term is negligibly important 
and the scattering cross section is approximately proportional 
to $\Delta\omega^{-2}$, where $\Delta\omega$
is the difference of the angular frequency of the incident radiation 
and that of the Ly$\alpha$ line center, $\omega_{Ly\alpha}$. This cross 
section given by the Lorentzian is symmetric 
with respect to the line center in frequency space.  However, 
the true scattering cross section 
deviates from $\Delta\omega^{-2}$  very far from the line center 
as the contributions increase from the other energy eigenstates
including both bound and continuum states.  
The zeroth order term $\sigma \propto (\omega_{Ly\alpha}/\Delta\omega)^{2}$ 
being symmetric with respect to the line center in frequency space,
the first order term $\Delta\omega/\omega_{Ly\alpha}$ for this deviation 
will contribute to the asymmetry of the scattering
cross section. Therefore, a very careful quantitative consideration
of this effect should be taken in order to obtain an accurate estimation 
of the true line center of high H~I column density absorption systems.

H~I absorption systems are easily found in typical quasar spectra and
are very useful to investigate the baryonic matter content and the 
large scale structure in the universe. The quasar H~I
absorption systems include the Ly$\alpha$ forest, Lyman limit
systems and damped Ly$\alpha$ systems, which are classified 
according to the H~I column density
$N_{HI}$.  A number of damped Ly$\alpha$ systems (DLA's) are known to 
exhibit H~I column density
in excess of $10^{21}{\rm\ cm^{-2}}$ (e.g. Turnshek \& Rao, 1998). 
Very high H~I column density media may be found in the early universe, 
when the reionization is not complete and the universe is only partially 
ionized by the first objects
(Gunn \& Peterson 1965, Scheuer 1965). In these high column H~I density media, 
the correct redshift of the absorbing system can be obtained after proper 
considerations of the asymmetric scattering cross section in frequency space.

In this paper, we calculate the asymmetric deviation of the scattering 
cross section of radiation around Ly$\alpha$ by atomic hydrogen and 
apply our results to high H~I column density scattering media 
that may be important in observational cosmology.

\section{Calculation}
\subsection{The Kramers-Heisenberg Formula}
We start with the Kramers-Heisenberg formula that is obtained from
the fully quantum mechanical second-order time dependent theory (e.g.
Sakurai 1969, Merzbacher 1970).
In terms of the matrix elements of the dipole operator $({e \bf x})$,
the differential cross section ${d\sigma\over d\Omega}$ for incident
radiation with angular frequency $\omega$ is given by
the Kramers-Heisenberg formula, which can be written as
\begin{eqnarray}
{d\sigma\over d\Omega}&=& {r_0^2 m_e^2 \over \hbar^2} \left|
\sum_n \omega\omega_{n1}\left({{({\bf x}\cdot \epsilon^{(\alpha')})_{1n}
({\bf x}\cdot \epsilon^{(\alpha)})_{n1}}\over{\omega_{n1}-\omega}} 
\right.\right.  \nonumber \\ 
&-& \left.  {{({\bf x}\cdot \epsilon^{(\alpha)})_{1n}
({\bf x}\cdot \epsilon^{(\alpha')})_{n1}}\over{\omega_{n1}+\omega}}
\right) \nonumber \\
&+&  \int_0^\infty dn'\omega\omega_{n'1}\left(
{{({\bf x}\cdot \epsilon^{(\alpha')})_{1n'}
({\bf x}\cdot \epsilon^{(\alpha)})_{n'1}}\over{\omega_{n'1}-\omega}} \right. 
\nonumber \\
&-& \left.\left. {{({\bf x}\cdot \epsilon^{(\alpha)})_{1n'}
({\bf x}\cdot \epsilon^{(\alpha')})_{n'1}}\over{\omega_{n'1}+\omega}}
\right)\right|^2. 
\end{eqnarray}
where $m_e$ is the electron mass,  $r_0=e^2/m_e c^2$ is the classical 
electron radius, 
$({\bf x}\cdot \epsilon^{(\alpha')})_{1n}$ represents the matrix element 
of the position operator between the ground $1s$ state and excited
$np$ state, and $\epsilon^{\alpha}, \epsilon^{\alpha'}$ are polarization 
vectors of incident and scattered radiation, respectively. 
Here, $\omega_{n1}$ is the angular frequency between the bound $np$ state 
and the ground $1s$ state, and $\omega_{n'1}$ is the angular frequency of the
transition between $1s$ and the continuum
$n'p$ state. In the atomic units adopted in this work, the bound $np$ state 
has an energy eigenvalue $E_n = -{1\over2 n^2}$ and correspondingly 
$\omega_{n1}=({1\over2})(1-1/n^2)$.  Similarly, 
$E_{n'}={1\over 2(n')^2}$ and $\omega_{n'1}=({1\over 2})(1+1/n'^2)$.

In the limit $\omega \rightarrow \infty$,  with the use of the closure relation
it can be shown that the Kramers-Heisenberg formula yields a constant scattering
cross section independent of the wavelength of incident radiation, which is 
the Thomson scattering cross section ${d\sigma/d\Omega} = r_0^2|\epsilon^{(\alpha)}
\cdot\epsilon^{(\alpha')}|^2$.
In the opposite limit where $\omega \ll \omega_{21} = \omega_{Ly\alpha}$,
another use of the completeness of the intermediate states gives
\begin{eqnarray}
{d\sigma\over d\Omega}
&=&\left({r_0 m_e\over\hbar}\right)^2\omega^4 \left| \sum_I
\left({1\over\omega_{I1}}\right)[({\bf x}\cdot \epsilon^{(\alpha')})_{1I} 
({\bf x}\cdot \epsilon^{(\alpha)})_{I1} 
\right. 
\nonumber \\
&+& \left.  ({\bf x}\cdot \epsilon^{(\alpha)})_{1I} 
({\bf x}\cdot \epsilon^{(\alpha')})_{I1}] 
\right|^2
\end{eqnarray}
which shows the well-known $\omega^4$ dependence of the Rayleigh 
scattering cross section in the low energy limit.

In order to investigate the behavior of the cross section around Ly$\alpha$, 
we introduce
\begin{equation}
\Delta\omega \equiv \omega-\omega_{21}
\end{equation}
and expand the Kramers-Heisenberg formula in terms 
of $\Delta\omega/\omega_{21}$.

In the first term inside the summation of Eq.~1 with $n=2$, we can write
\begin{equation}
{\omega\omega_{21} \over \omega_{21}-\omega}=-\omega_{21}
\left(1+{\omega_{21}\over \Delta\omega} \right)
\end{equation}
In the same term for $n\ge 3$, we may write
\begin{equation}
{\omega \over \omega_{n1}-\omega} =
{\omega_{21}\over\omega_{n1}-\omega_{21}}
\left[ 1 +{\omega_{n1}\over \omega_{21}}
\sum_{k=1}^\infty \left( {\Delta\omega\over \omega_{n1}-\omega_{21}} \right)^k \right].
\end{equation}
In a similar way, for $n\ge 2$, the second term in the summation in Eq.~1 
can be written as
\begin{equation}
{\omega \over \omega_{n1}+\omega} = 
{\omega_{21}\over \omega_{n1}+\omega_{21}}
\left[ 1 -{\omega_{n1}\over \omega_{21}}
\sum_{k=1}^\infty \left( {-\Delta\omega\over\omega_{n1}+\omega_{21}} \right)^k
\right].
\end{equation}
Similar expressions for continuum states are obtained 
in a straightforward manner.

Making use of the Wigner-Eckart theorem we may separate the
matrix elements of the position operator into the angular part and
the radial part, where the radial part is given by the reduced matrix elements
$<f\parallel x\parallel i>$.  (e.g. Merzbacher 1970). The angular part gives the
same dipole type angular distribution as that for Thomson scattering or 
classical Rayleigh scattering (e.g. Ahn, Lee \& Lee 2000, Lee 1997, 
Schmid 1989).

Substituting Eqs~5,6 into Eq.~1, we obtain
\begin{eqnarray}
\sigma &=& \sigma_T \left({\omega_{21}\over\Delta\omega}\right)^2
\left| \omega_{21}<x>_{12}^2 +\omega_{21}<x>_{12}^2 
\left({\Delta\omega\over \omega_{21}}\right) \right. \nonumber \\
&-& \sum_{n=3}^\infty
{\Delta\omega\omega_{n1}\over\omega_{n1}-\omega_{21}}
\left[ 1 +{\omega_{n1}\over \omega_{21}}
\sum_{k=1}^\infty \left( {\Delta\omega\over \omega_{n1}-\omega_{21}} \right)^k \right]
<x>_{n1}^2 
\nonumber \\
&+& \sum_{n=2}^\infty
{\Delta\omega\omega_{n1}\over \omega_{n1}+\omega_{21}}
\left[ 1 -{\omega_{n1}\over \omega_{21}}
\sum_{k=1}^\infty \left( {-\Delta\omega\over\omega_{n1}+\omega_{21}} \right)^k
\right] <x>_{n1}^2  \nonumber \\
&-& \int_0^\infty dn'
{\Delta\omega\omega_{n'1}\over\omega_{n'1}-\omega_{21}}
\left[ 1 +{\omega_{n'1}\over \omega_{21}}
\sum_{k=1}^\infty \left( {\Delta\omega\over \omega_{n'1}-\omega_{21}} \right)^k \right]
<x>_{n'1}^2 
\nonumber \\
&+& \left. \int_0^\infty dn'
{\Delta\omega\omega_{n'1}\over \omega_{n'1}+\omega_{21}}
\left[ 1 -{\omega_{n'1}\over \omega_{21}}
\sum_{k=1}^\infty \left( {-\Delta\omega\over\omega_{n'1}+\omega_{21}} \right)^k
\right] <x>_{n'1}^2 \right|^2
\end{eqnarray}
where the atomic units have been used and angular integration has been 
performed. Here, the Thomson scattering cross section 
$\sigma_T = 8\pi r_0^2/3=0.665\times 10^{-24}{\rm\ cm^2}$.  
The matrix elements of the dipole operators $<x>_{n1}, <x>_{n',1}$ are 
easily found in textbooks on quantum mechanics (e.g. Berestetski, 
Lifshitz \& Pitaevskii 1971, Saslow \& Mills 1969, Bethe \& Salpeter 1967). 
The matrix elements are given by
\begin{equation}
<x>_{n1}=<1s| x | np> =\left[ {2^{8}n^7 (n-1)^{2n-5}}\over
{3(n+1)^{2n+5}} \right]^{1\over2} a_0 
\end{equation}
for the bound states.  Here, the Bohr radius $a_0=\hbar^2/ me^2$ is set 
equal to 1 in atomic units.
For the continuum states, the corresponding values are given by
\begin{equation}
<x>_{n'1} = <1s| x| n'p>
=\left[ {2^{8}(n')^7 \exp[-4n'\tan^{-1}(1/n')]}\over
{3[(n')^2+1]^{5}[1-\exp(-2\pi n')]} \right]^{1\over2} 
\end{equation}
where we also adopt the atomic units.

Eq.~7 can be written as
\begin{equation}
\sigma = \sigma_T\left( {\omega_{21}\over\Delta\omega} \right)^2
\left|A_0+A_1\left({\Delta\omega\over\omega_{21}}\right)  
 +A_2\left({\Delta\omega\over\omega_{21}}\right)^2 
  +\cdots \right|^2 
\end{equation}
where the coefficients $A_k$ are obtained through the following relations
\begin{eqnarray}
A_0 &=& \omega_{21}<x>_{12}^2 ={1\over2} f_{12} =0.2081\nonumber \\
A_1 &=& {3\over2}\omega_{21}<x>_{12}^2 
-{1\over2}\sum_{n=3}^\infty {\omega_{21}^2\omega_{n1}
\over\omega_{n1}^2-\omega_{21}^2}<x>_{1n}^2 =  \nonumber \\
&-&{1\over2}\int_0^\infty dn' {\omega_{21}^2\omega_{n'1}
\over\omega_{n'1}^2-\omega_{21}^2}<x>_{n1'}^2 =-0.1865
\nonumber \\
A_k &=&\left({-1\over 2}\right)^k\omega_{21}<x>_{12}^2
\nonumber \\
&-&\sum_{n=3}^\infty {\omega_{n1}^2\over\omega_{21}}
\left[
\left({\omega_{21}\over\omega_{n1}-\omega_{21}}\right)^k
-\left({-\omega_{21}\over\omega_{n1}+\omega_{21}}\right)^k
\right]<x>_{n1}^2 \nonumber \\
&-&\int_{0}^\infty dn' {\omega_{n'1}^2\over\omega_{21}}
\left[
\left({\omega_{21}\over\omega_{n'1}-\omega_{21}}\right)^k
-\left({-\omega_{21}\over\omega_{n'1}+\omega_{21}}\right)^k
\right]<x>_{n'1}^2. 
\end{eqnarray}
Here, $f_{12}=0.4162$ is the oscillator strength for the Ly$\alpha$ transition.

The computation of the coefficients can be done in a straightforward way, from
which we obtain a list of these coefficients up to $k=5$ 
\begin{eqnarray}
A_1/A_0 & = & -0.8961  \nonumber \\
A_2/A_0 & = & -1.222\times 10^1  \nonumber \\
A_3/A_0 & = & -5.252\times 10^1   \nonumber \\
A_4/A_0 & = & -2.438\times 10^2   \nonumber \\
A_5/A_0 & = & -1.210\times 10^3.   
\end{eqnarray}

Therefore, up to the first order in $\Delta\omega/\omega_{21}$, the scattering 
cross section of radiation around Ly$\alpha$
\begin{eqnarray}
\sigma(\omega) &\simeq& \sigma_T\left({f_{12}\over 2}\right)^2 
\left({\omega_{21}\over\Delta\omega} -0.8961
\right)^2 \nonumber \\
&\simeq& \sigma_T \left({f_{12}\over2}\right)^2 \left({\omega_{21}\over
\Delta\omega}\right)^2
\left(1-1.792{\Delta\omega\over\omega_{21}}\right)
\end{eqnarray}

In Fig.~1, we show the scattering cross section multiplied by a column 
density $N_{HI}=4\times 10^{22}{\rm\ cm^{-2}}$, that is the scattering
optical depth of a neutral medium with this column density. 
The solid line represents
the result obtained from the Kramers-Heisenberg formula, the dotted line
depicts the first order 
approximation to it obtained in Eq.~13 and the dashed line shows 
the zeroth approximation 
proportional to $\Delta\omega^{-2}$ obtained
from the Lorentzian by ignoring the damping part. The inclusion 
of the first order term implies the introduction of asymmetry of 
the scattering cross section with
respect to the line center in frequency space. As is apparent in Fig.~1, the 
scattering cross section is smaller than the Lorentzian blueward of 
the line center, whereas it is larger in the opposite part.

\subsection{Applications to High $N_{HI}$ Media}

In Fig.~2 we show the absorption profile $e^{-\tau(\lambda)}$ 
that is expected from a damped Ly$\alpha$ system with a column 
density $N_{HI}=5\times 10^{21}{\rm\ cm^{-2}}$.  The thick solid line
represents the result obtained using the first-order corrected formula and
the dotted line is the absorption profile obtained using the Voigt profile 
by shifting the center wavelength by an amount 
$\Delta\lambda =0.187{\rm\ \AA}$ to the blue part.
The wavelength shift by this amount is determined using the least square 
method that best fits the thick solid line.  Because in the observational 
data it is difficult to obtain the exact continuum level,
we minimize the difference $|e^{-\tau_L(\lambda)}-e^{-\tau_{LC}(\lambda)}|^2$
in the wavelength interval between 1180 \AA\ and 1250 \AA, 
where the suppression of the continuum is larger than 20 percent.

Therefore, if we apply the Voigt profile fitting process to observational data
for an H~I scattering system with $N_{HI}=5\times 10^{21}{\rm\ cm^{-2}}$, 
we may underestimate the redshift of the H~I system by an amount of $\Delta z
=0.187/1215.67 = 1.54\times 10^{-4}$ or in terms of the velocity scale
$\Delta v= 39{\rm\ km\ s^{-1}}$. 

Rao \& Turnshek (2000) investigated the damped Ly$\alpha$ absorption 
system along the
line of sight to the quasar PKS 1127-145, which was best fitted with the Voigt 
profile with $z=0.3127$ (see also Rao et al. 2003). Hence, the
first order correction from our quantum mechanical consideration may affect
the fourth decimal place of $z$, which is not critically important 
considering the accuracy of $\Delta z=0.001$ quoted by the authors. 
Among DLA's so far investigated, the highest H~I column density 
is $\sim N_{|HI}=5\times 10^{21}{\rm\ cm^{-2}}$ (Rao \& Turnshek 2000). 

According to Braun (1997), gas rich galaxies may contain a medium 
with $N_{HI}=4\times 10^{22} {\rm\ cm^{-2}}$. In Fig.~3 we show
the scattering optical depth for this H~I column density.
For a higher column density system with 
$N_{HI}=4\times10^{22}{\rm\ cm^{-2}}$, we obtain
$\Delta\lambda = 1.17{\rm\ \AA}$, which may result in underestimation 
of the redshift by $\Delta z = 0.962\times 10^{-3}$. This amounts 
to a velocity shift of 
$\Delta v=289{\rm\ km\ s^{-1}}$, and may affect significantly 
the exact redshift determination of the absorbing system.

\section{Discussion}

Gavrila (1967) provided a quantum mechanical computation of the Rayleigh 
scattering cross section using the Green function in momentum space.
Ferland (2001) obtained the power fit to the result of Gavrila (1967) and 
incorporated it in his photoionization code `Cloudy,' in which the formula 
is useful for radiation with
$\lambda >1410{\rm\ \AA}$ and for radiation nearer the Ly$\alpha$ line center
Voigt function is proposed to be used. 
In the low energy limit where $\omega \ll \omega_{Ly\alpha}$, the scattering 
cross section tends to be proportional to $\omega^4$, which is also 
expected from classical physics.  This approximation is quite useful in
dealing with basic radiative transfer, because the Voigt function 
(or equivalently the Lorentzian
function) well approximates the true scattering cross section. 

Therefore, in this work, we emphasize the asymmetric deviation of the true 
scattering cross section from the Lorentzian in frequency space, 
which may result in underestimation of the exact wavelength 
of the Ly$\alpha$ line center.  Careful comparisons of redshifts 
determined using other lines may provide important clues to the kinematics 
of the absorbing systems.  In particular, Prochaska \& Wolfe (1997, 1998) 
proposed that the absorption profiles observed in many DLA's are consistent 
with the hypothesis that they are formed in a rotating disk.  
The apparent line shift due to the asymmetric deviation 
around Ly$\alpha$ may affect the kinematic analysis of rotating galaxies. 

Scattering of Ly$\alpha$ photons is also important in search for the first 
objects that are responsible for the reionization of the universe 
(Gunn \& Peterson 1965, Scheuer 1965). In several high red shift
quasars with $z>6.2$ the presence of the Gunn-Peterson trough has 
been reported blueward of Ly$\alpha$ (Becker et al. 2001, Fan et al. 2003).  
The Gunn-Peterson optical depth will be important in the red wing part of
Ly$\alpha$ (e.g. Miralda-Escud\'e 1998).
In Eq.~(13) the correction from the Lorentzian cross section is about 1 \%
when we consider a width of the damped profile of about 
$2000{\rm\ km\ s^{-1}}$, for the characteristic expected value of the
intergalactic medium (IGM) density at around $z=6$. Hence, the change
in the cross section is equivalent to a change in the redshift of the
Gunn-Peterson trough of about 1 \% of th ewidth of the damped profile
or $20{\rm\ km\ s^{-1}}$, when only the first order term in $\Delta\omega
/\omega_{21}$ is considered.

Another high column density media may be found in the atmospheres of 
supergiants, in which classical Rayleigh scattering is important 
(e.g. Isliker Nussbaumer \& Vogel 1989). 
Also about a half of the symbiotic stars believed to consist of a heavily mass 
losing giant and a 
white dwarf exhibit Raman scattered O~VI features around 6830 \AA\ and 
7088 \AA\ (Schmid 1989, Nussbaumer, Schmid, \& Vogel 1989). These features are 
originated from
the resonance doublet O~VI 1032, 1038 line photons that are scattered
in a neutral region with $N_{HI}\sim 10^{22}{\rm\ cm^{-2}}$. 
However, due to resonant scattering by
neutral hydrogen in interstellar space, it will be difficult to accurately
determine the asymmetry of the scattering cross section.

\acknowledgments
The author is very grateful to an anonymous referee for useful comments.
This work is a result of research activities of the Astrophysical Research 
Center for the Structure and Evolution of the Cosmos (ARCSEC) funded by 
the Korea Science
and Engineering Foundation.

\clearpage

\begin{figure}
\plotone{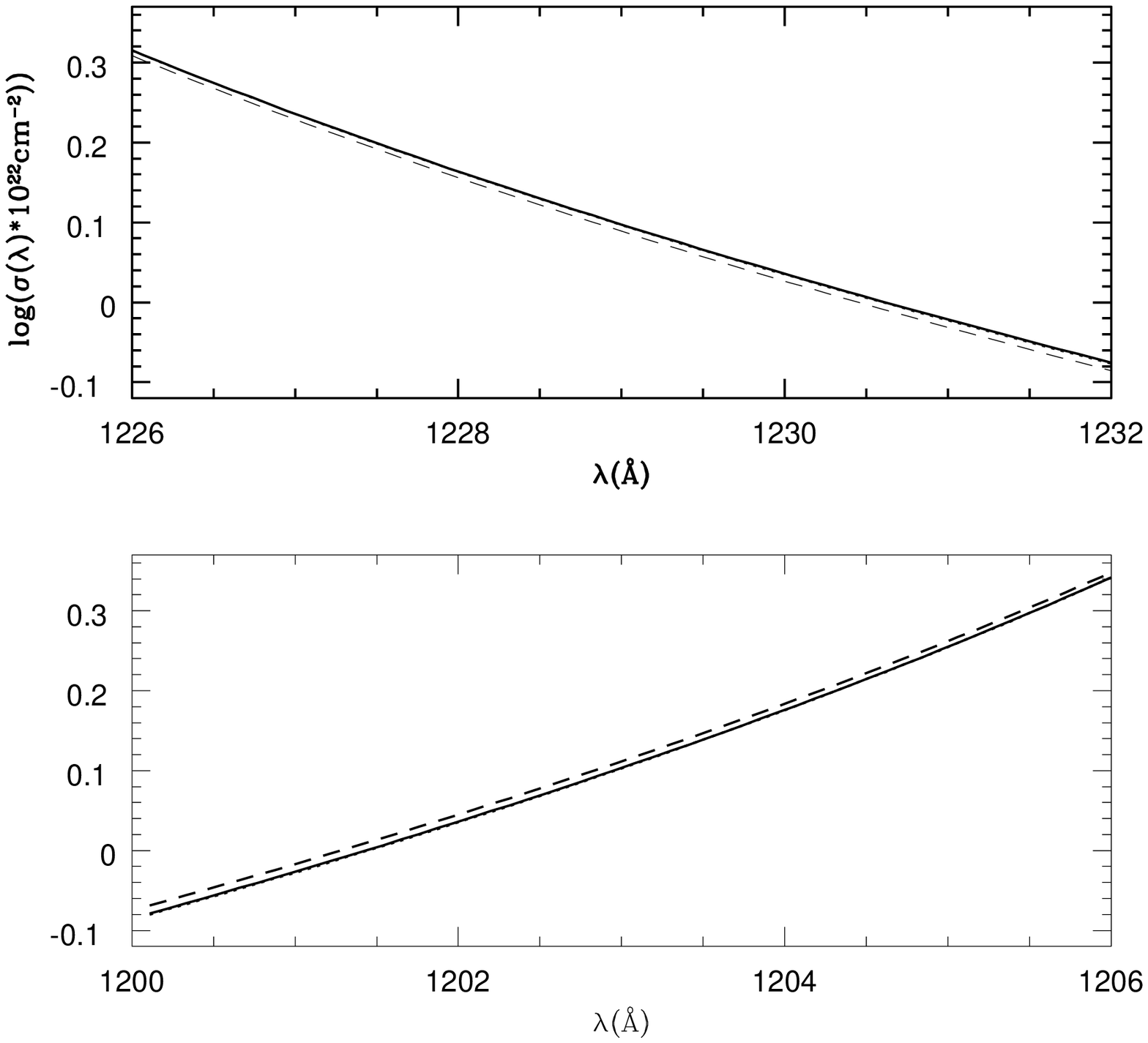} \caption{
The scattering cross section around Ly$\alpha$. The result obtained from 
the Kramers-Heisenberg formula is shown by the solid line, the first order 
approximation to it obtained in Eq.~13 and the zeroth approximation 
that is proportional to $\Delta\omega^{-2}$ obtained
from the Lorentzian by ignoring the damping part. 
The scattering cross section is smaller than the Lorentzian blueward of 
the line center, whereas it is larger in the opposite part.
\label{fig1}}
\end{figure}

\begin{figure}
\plotone{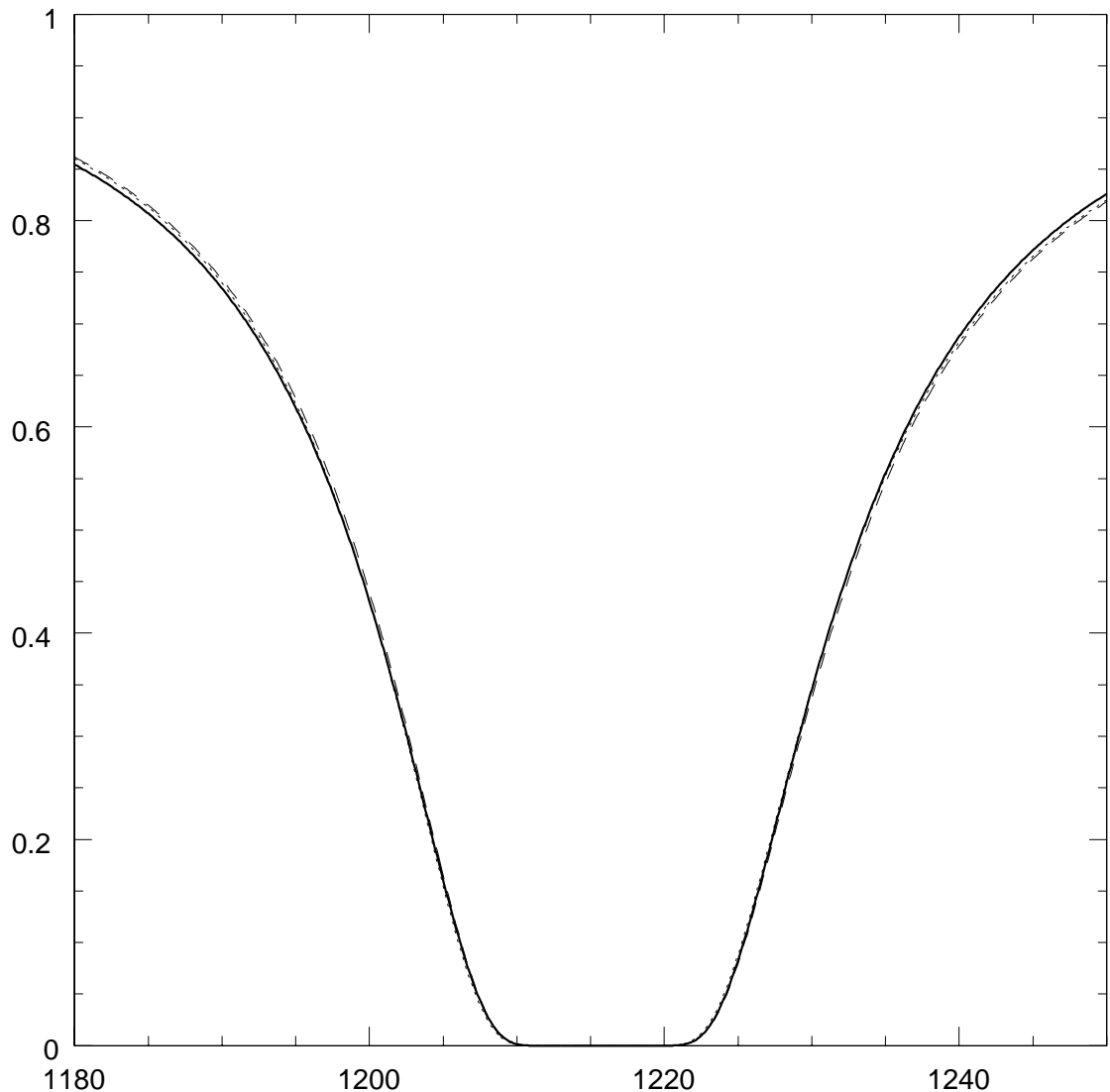} \caption{
The absorption profile $e^{-\tau(\lambda)}$ 
expected from a damped Ly$\alpha$ system with a column 
density $N_{HI}=5\times 10^{21}{\rm\ cm^{-2}}$.  The thick solid line
represents the result obtained using the first-order corrected formula and
the dotted line is the absorption profile obtained using the Voigt profile 
by shifting the center wavelength by an amount 
$\Delta\lambda =0.187{\rm\ \AA}$ to the blue part.
The wavelength shift by this amount is determined using the least square 
method that best fits the thick solid line.  
\label{fig2}}
\end{figure}

\begin{figure}
\plotone{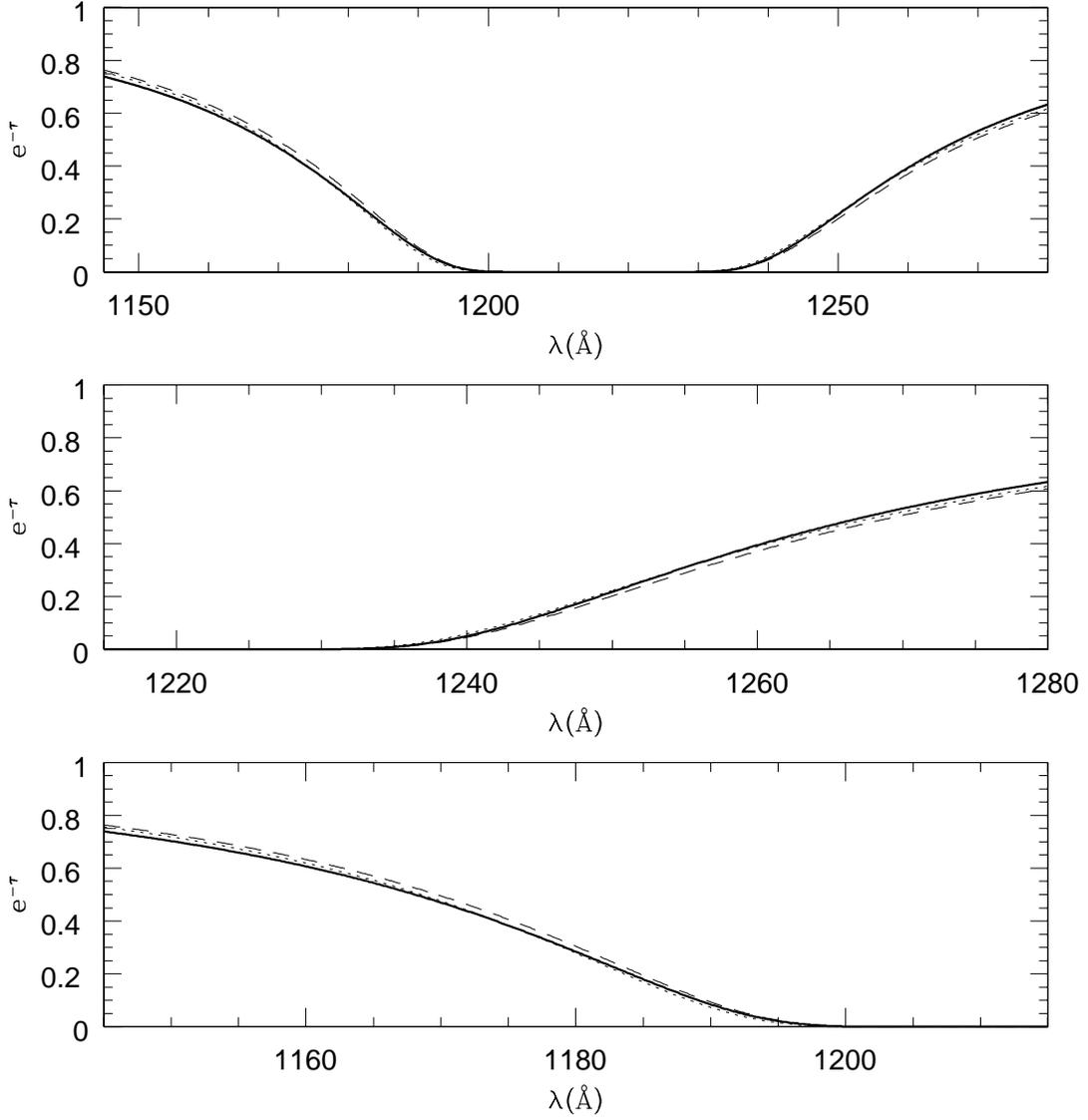} \caption{
The absorption profile $e^{-\tau(\lambda)}$ 
expected from a damped Ly$\alpha$ system with a column 
density $N_{HI}=4\times 10^{22}{\rm\ cm^{-2}}$. The thick solid line
represents the first-order correction. The best-fit wavelength shift
from the Lorentzian fit is
$\Delta\lambda = 1.17{\rm\ \AA}$, which may result in underestimation 
of the redshift by $\Delta z = 0.962\times 10^{-3}$.
\label{fig3}}
\end{figure}

\end{document}